\documentclass[a4paper,12pt]{article}
\usepackage{amsmath,amsfonts,amssymb,epsfig,subfigure,color}

\usepackage{caption}
\usepackage[percent]{overpic}
\usepackage{amssymb}
\usepackage{amsmath, epsfig, color, subfigure,bm, graphics}
\usepackage{fullpage}
\usepackage{color}
\newcommand{\be}{\begin{equation}}
\newcommand{\en}{\end{equation}}
\renewcommand{\vec}[1]{\boldsymbol{#1}}

\numberwithin{equation}{section}

\begin{document}

\numberwithin{equation}{section}

\title{Extreme softness of brain matter in simple shear\footnote{
\emph{dedicated to Martine Ben Amar in esteem and friendship}}}

\author{M. Destrade$^{1,2}$, M.D. Gilchrist$^2$, \\
J.G. Murphy$^{1,3}$, B. Rashid$^{2,4}$, G. Saccomandi$^{1,5}$\\[12pt]
$^1$School of Mathematics, Statistics and Applied Mathematics,\\
National University of Ireland Galway,\\ University Road, Galway, Ireland;\\[12pt]
$^2$School of Mechanical and Materials Engineering, \\
University College Dublin, Belfield, Dublin 4, Ireland;\\[12pt]
$^3$Centre for Medical Engineering Research, \\
Dublin City University, Glasnevin, Dublin 9, Ireland;\\[12pt]
$^4$Research and Development, \\
Pakistan Ordnance Factories,
47040 Wah Cantt, Pakistan;\\[12pt]
$^5$Dipartimento di Ingegneria,\\ 
Universit\`a di Perugia, Via G. Duranti, Perugia 06125, Italy.}

\date{}

\maketitle


\begin{abstract}

We show that porcine brain matter can be modelled accurately as a very soft rubber-like material using the Mooney-Rivlin strain energy function, up to strains as high as 60\%. 
This result followed from simple shear experiments performed on small rectangular fresh samples ($2.5$ cm$^3$ and $1.1$ cm$^3$) at quasi-static strain rates. 
They revealed a linear shear stress--shear strain relationship ($R^2> 0.97 $), characteristic of Mooney-Rivlin materials at large strains. 
We found that porcine brain matter is about 30 times less resistant to shear forces than a silicone gel.
We also verified experimentally that brain matter exhibits the positive Poynting effect of nonlinear elasticity, and numerically that the stress and strain fields remain mostly homogeneous throughout the thickness of the samples in simple shear. 

\end{abstract}


\noindent
\textbf{keywords:}
mechanics of brain matter;
simple shear;
constitutive modelling;
experimental testing;
simulations;
Poynting effect


\section{Introduction
\label{Introduction}}


Modelling the mechanical properties of brain matter is quite a straightforward process once it is accepted that a small enough sample can be considered to be isotropic, homogeneous, nonlinear and viscoelastic.
However, testing its mechanical behaviour in order to evaluate its constitutive parameters is not a straightforward matter at all, and is fraught with potential modelling mistakes and experimental pitfalls. 

For example, standard \emph{tensile tests} cannot  be performed properly (i.e. cannot follow standard practice, e.g. \cite{BS-tension}) on brain matter, because it fractures easily in that regime. 
As a consequence of that brittleness, it is not practical to cut dog-bone shaped specimens and one must then resort to using cylindrical or prismatic samples, to be glued at their extremities to the cross-heads of a tensile machine, for \emph{simple tension} or \emph{simple compression} tests. 
However, because the faces are glued, end effects intervene early in the deformation, see  \cite{RaDG12}.
They make the deformation field strongly inhomogeneous with the consequence that the stress-strain relationship becomes impossible to determine analytically.

\begin{figure}[!t] 
  \begin{center}
\subfigure{\reflectbox{ \includegraphics[width=0.47\columnwidth] {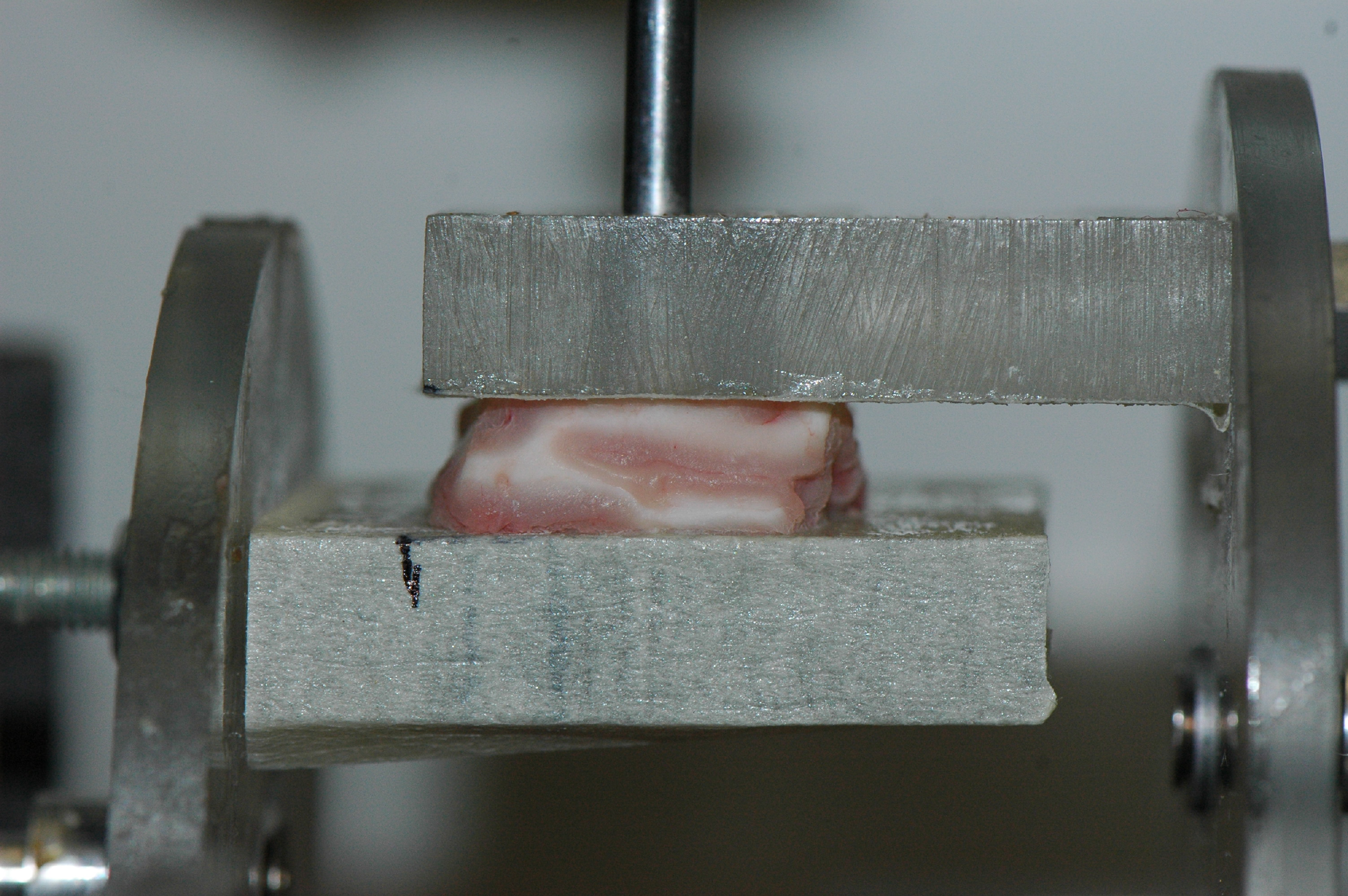} }}
\subfigure{\reflectbox{ \includegraphics[width=0.47\columnwidth] {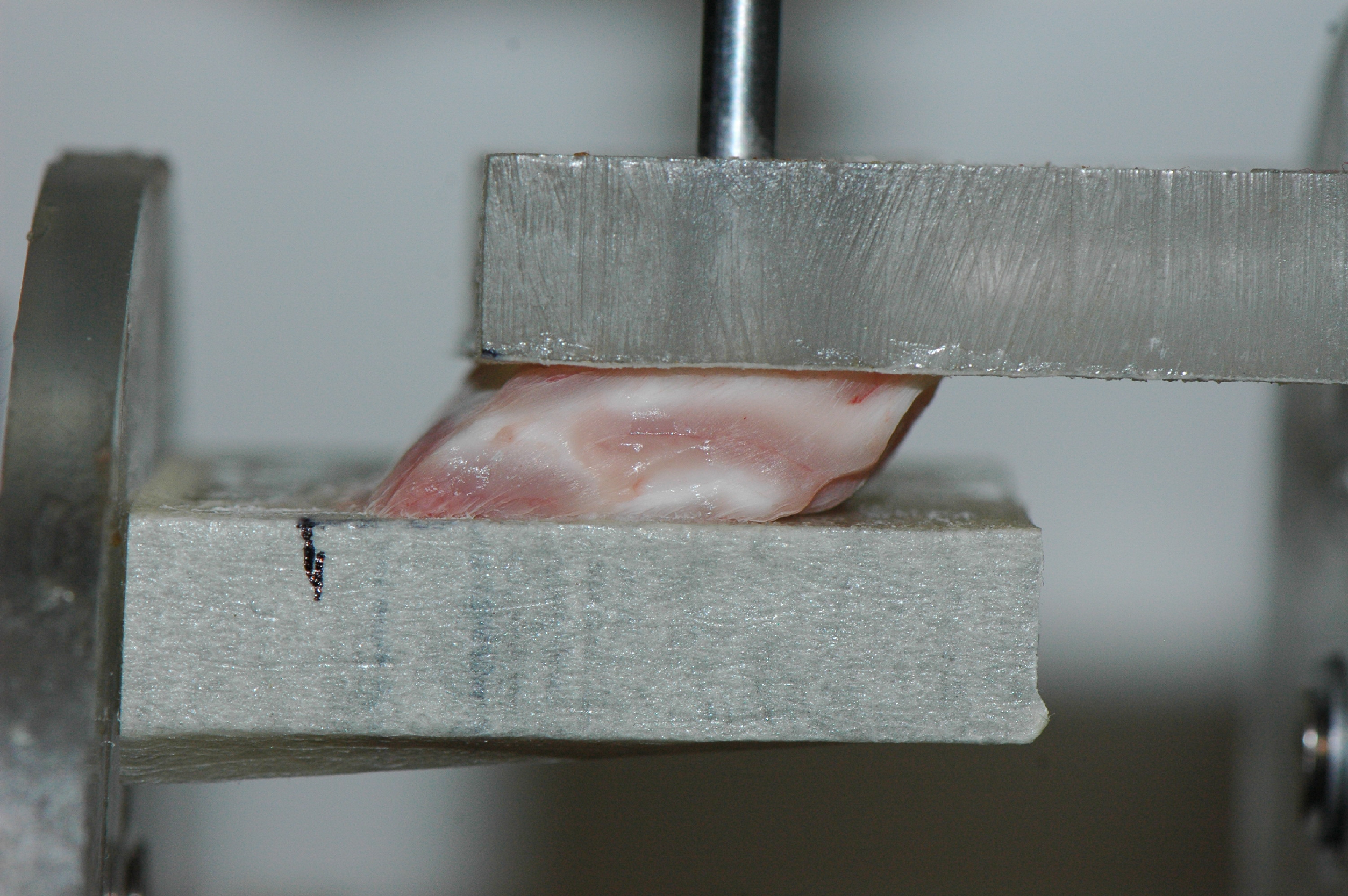} }}
\end{center}
\caption{
\small
Simple shear testing of porcine brain matter. Here the sample is a cuboid with dimensions $19 \times 19 \times 7$ mm, and has two opposite faces glued to parallel platens, \color{black} the top one fixed and the bottom one mobile. \color{black}
The picture on the right was taken after a bottom platen displacement of 7 mm, corresponding to an amount of shear $K=1$, an angle of shear of $45^\circ$ and a maximal stretch of 62\%.
The edge effects are very localised and most of the sample seems to have deformed homogeneously.}
\label{fig:shear}
\end{figure}

Another  standard testing protocol exists \cite{ISO} which has received a lot less  attention for brain matter, namely the \emph{simple shear} test. 
This is a most illuminating homogeneous deformation, which brain matter seems to be able to withstand quite well, see Fig.\ref{fig:shear} where the amount of shear is $K=1.0$, corresponding to a maximal stretch of $K/2 + \sqrt{1+K^2/4} = 1.618$, i.e. an extension of 62\%.

In this paper we show that careful modelling and experimental data acquisition lead to an almost complete characterisation of brain matter as an incompressible isotropic nonlinear elastic material. 
Hence, we show experimentally that forces normal to the platens develop during large simple shear, and this effect allows us to rule out the entire class of generalised neo-Hookean solids. 
Also, we obtain a \emph{linear} shear-stress/amount-of-shear relationship, a property characteristic of the Mooney-Rivlin material, with strain energy density
\be
W = C_1(I_1-3) + C_2(I_2-3),
\label{MR}
\en
where $C_1$, $C_2$ are constants and $I_1$, $I_2$ are the first two principal invariants of the left Cauchy-Green deformation tensors.

The conclusion is that it is thus sufficient to determine only two material parameters in order to fully characterise the (quasi-static) mechanical behaviour of brain matter, up to at least  60\%, which is more than adequate for practical purposes, including the study of diffuse axonal injury (DAI), which is believed to occur at macroscopic shear strains of approximately 10\%--50\% (see e.g. \cite{BaMe00, BaBP06}).

Other  advances put forth by this work are the {exhibition of the normal force effects} generated by simple shear in brain matter, the numerical verification that the experimental protocol does indeed produce almost homogeneous fields, and the comparison of the stiffness of brain matter with that of a silicone gel at quasi-static speeds.
We found that at quasi-static strain rates, the former is about 30 times softer in shear than the latter.


\section{Material and methods \label{Material and methods}}


\subsection{Analytical modelling of simple shear}

\begin{figure}[!t] 
  \begin{center}
\epsfig{figure=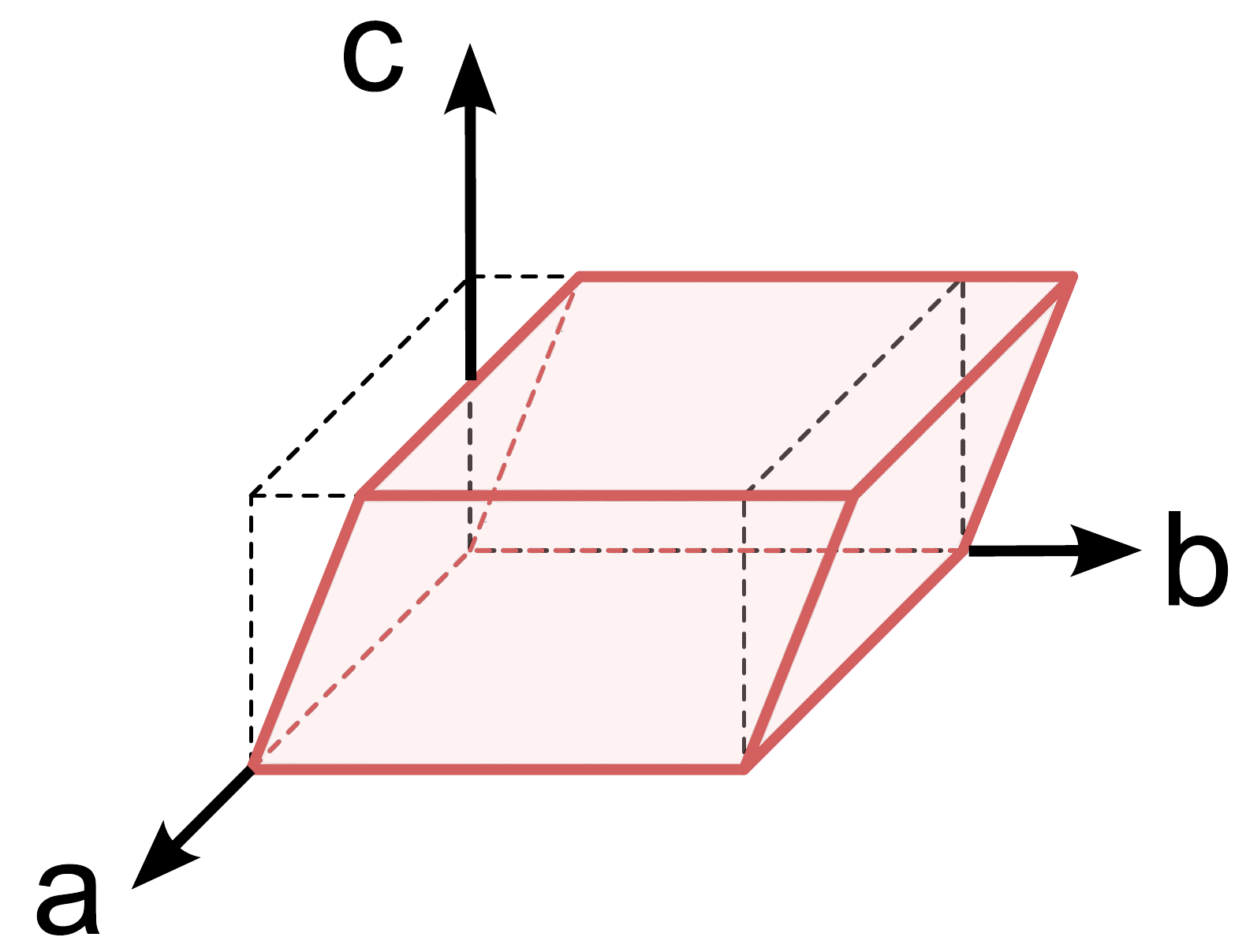, width=0.5\textwidth}
\end{center}
\caption{
\small
Simple shear deformation. Black lines: sample in the undeformed configuration; Red lines: deformed sample.}
\label{fig:shear_modelling}
\end{figure}

Let us consider a cuboid made of a homogeneous, isotropic, incompressible, nonlinear elastic solid subjected to a simple shear deformation such that: 
\be
x_1=X_1+K X_2,\qquad
x_2=X_2,\qquad
x_3=X_3,
\en
where $x_i$ denotes the coordinates in the current configuration, $X_i$ the coordinates in the reference configuration, and $K$ is the amount of shear, see Fig.\ref{fig:shear_modelling}.
 
As established by Rivlin \cite{Rivl49}, the following Cauchy stress components $\sigma_{ij}$ maintain the block in a static state of simple shear, 
\begin{align}
\label{sigma}
& \sigma_{11}=2K^2\dfrac{\partial W}{\partial I_1},  
&& \sigma_{22}=-2K^2\dfrac{\partial W}{\partial I_2}, 
&& \sigma_{12}=2K\left(\dfrac{\partial W}{\partial I_1}+\dfrac{\partial W}{\partial I_2}\right), \notag \\
 & \sigma_{13} =0, && \sigma_{23} = 0, && \sigma_{33} = 0,
\end{align}
where $W$ is the strain energy function. 

When a solid is modelled as a generalised neo-Hookean material, for which $W=W(I_1)$ only, the formula \eqref{sigma}$_2$ predicts that $\sigma_{22}$ should be zero. This observation will provide a universal check on the validity of such an assumption \cite{HoSm12, Peppe}.

When a solid is modelled by the Mooney-Rivlin model \eqref{MR} , these formulas give 
\be
\label{sigma_MR}
\sigma_{12}=2(C_1+C_2) K
\qquad
\sigma_{22} = - 2 C_2 K^2.
\en
In other words, a material modelled by the Mooney-Rivlin model is expected to have a linear shear response and a non-zero normal force effect. 
This latter nonlinear effect is the so-called \emph{Poynting effect}: when $C_2>0$, it predicts that the sheared sample should expand in the direction normal to the gliding plane.

\subsection{Sample preparation}

Tests were performed on two fresh porcine brains coming from $6$ month old pigs collected about $12$ hours after death from a local slaughter house. The brains were kept in a saline solution at $4^{\circ}$C to $5^{\circ}$C. Time between animals' death and testing was inferior to $5$ hours at most. 

Then two cuboids were excised with the help of a rectangular cutter-die from the cerebrum region of the two different brains containing mixed and grey matter. 
Two different thicknesses were obtained by removing excessive matter from the cutter-die by using a surgical scalpel blade. After cutting, a measure of the dimensions of the cuboids gave \color{black} $19 \times19 \times 7$ mm$=2.5$ cm$^3$ for Sample $A$ and $19 \times 19 \times 3$ mm$ =1.1$ cm$^3$ for Sample $B$. \color{black}

Another cuboid was also prepared, made of Sylgard gel with the same dimensions of Sample $A$.

\subsection{Testing set-up}
The top and bottom large sides of a specimen were glued between parallel platens with a thin layer of Cyanoacrylate (Low viscosity Z105880-1EA, Sigma-Aldrich, Wicklow, Ireland). 
The use of machined spacers between the two platens avoided the overstressing of the brain samples as well as it ensured a good parallelism. 
This assembly was then mounted on a custom-made apparatus, whose full details are given in \cite{Rashid}. 

The top platen was fixed and the bottom platen was mobile in the horizontal direction only, at the slow speed of 0.257 mm/min. 
The displacement of the moving platen was measured via a Linear Variable Displacement Transducer (LVDT), and 500 grams load-cells (GSO Series, Transducer Techniques, Temecula, California, USA) attached to each platen allowed to determine the required instantaneous shear force $F$. 
Once divided by the area $a$ of the face glued to the platen it gave the Cauchy shear stress component $\sigma_{12}=F / a$. 

At this juncture, it is worth noticing that simple shear is a plane stress--plane strain deformation, so that no change of area occurs for out-of-place surfaces and $a=a_0$, the initial area of $19 \times 19 = 361$ mm$^2$ for the samples (hence, Donnelly and Medige \cite{DoMe97} are in error when they write that $a_0=a\sqrt{1+K^2}$).

\subsection{Tests}

We obtained quasi-static simple shear data by recording $7$ measurements of the force on the samples with thickness $d = 7.0$ mm (Sylgard gel and Sample $A$ of brain matter), and $10$ measurements on the sample with thickness $d = 3.0$ mm (Sample $B$ of brain matter). 
The amount of shear was varied from $K=0$ (displacement = 0 mm) to $K=1$ (displacement $= d$ mm), leaving the samples to rest for $70$ s between consecutive increments, ensuring quasi-static rates.

We then drilled a $6.0$ mm diameter hole in the fixed platen, and re-started simple shear tests again in order to observe visually whether the upper face of the samples had a tendency to expand, contract, or remain flat. 

\subsection{Numerical simulation}

Parallel to the experiments, we conducted a numerical simulation of large simple shear experiments, using the commercial Finite Element Method code ABAQUS.

 The specimen geometry was developed as two-dimensional: the length of the specimen was taken as $19.0$ mm, its height as $3.00$ mm. The top platen had a centred hole of $6.00$ mm diameter, and was otherwise rigidly attached with the specimen and constrained in all directions. The lower edge of the specimen was displaced at a velocity of $3.0$ mm/s ($1$/s strain rate). A total of $553$ CPS$4$R elements ($4$ node bilinear plane stress quadrilateral, reduced integration hourglass control) were used for the brain part. 
 
 The material density was $1040$ kg/m$^3$ (as established by weighing the samples).
The Mooney-Rivlin parameters used were $C_1=7.5$ Pa and $C_2=$ 66 Pa; these particular values were chosen for illustrative purposes, but such that $2(C_1+C_2)=165$ Pa, in line with our experimental results for brain matter, see next section. 

\section{Results}

\subsection{Stress-strain profiles}
\begin{figure}
\subfigure{\epsfig{figure=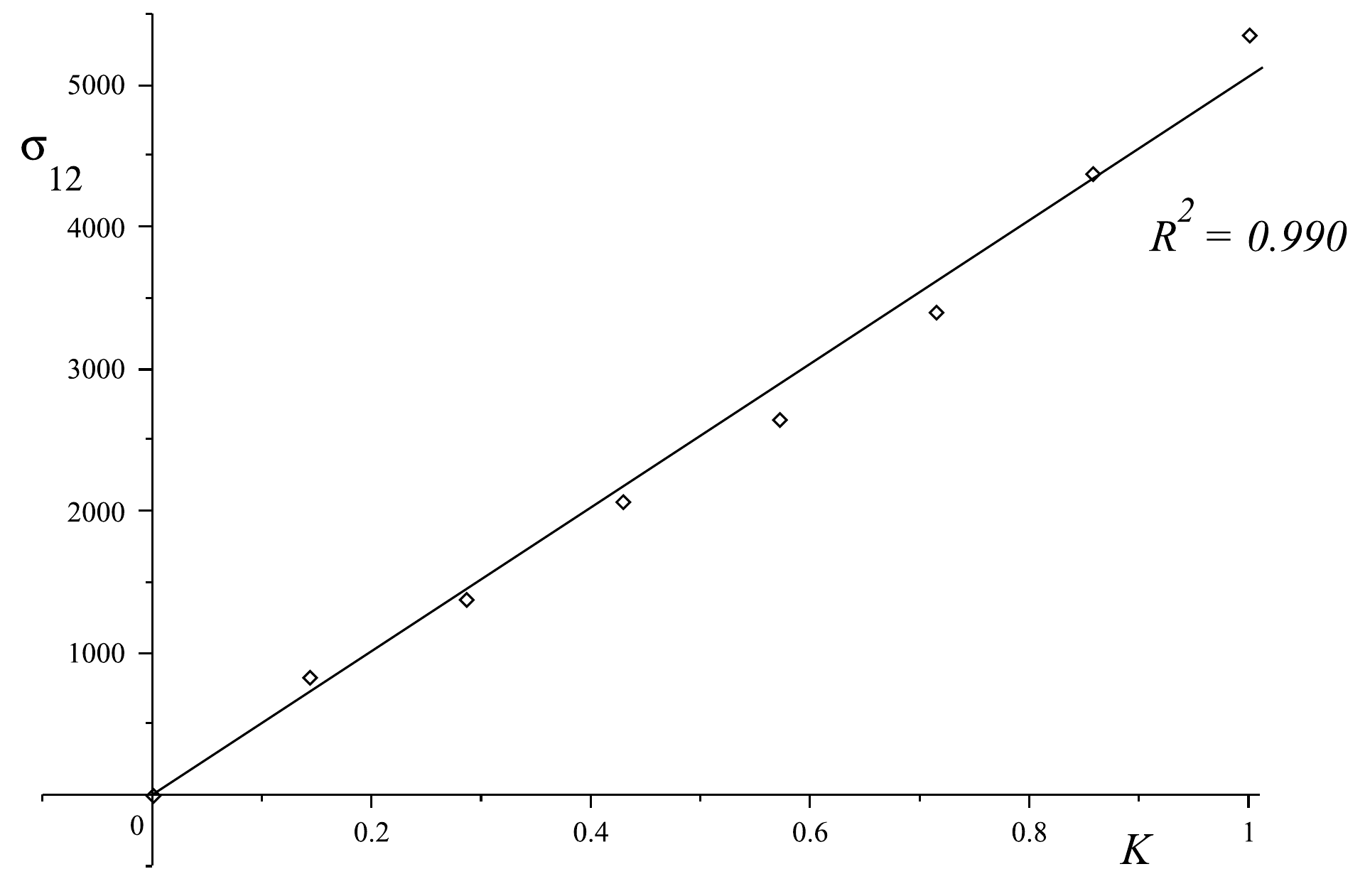, width=0.6\textwidth}}\\
\subfigure{\epsfig{figure=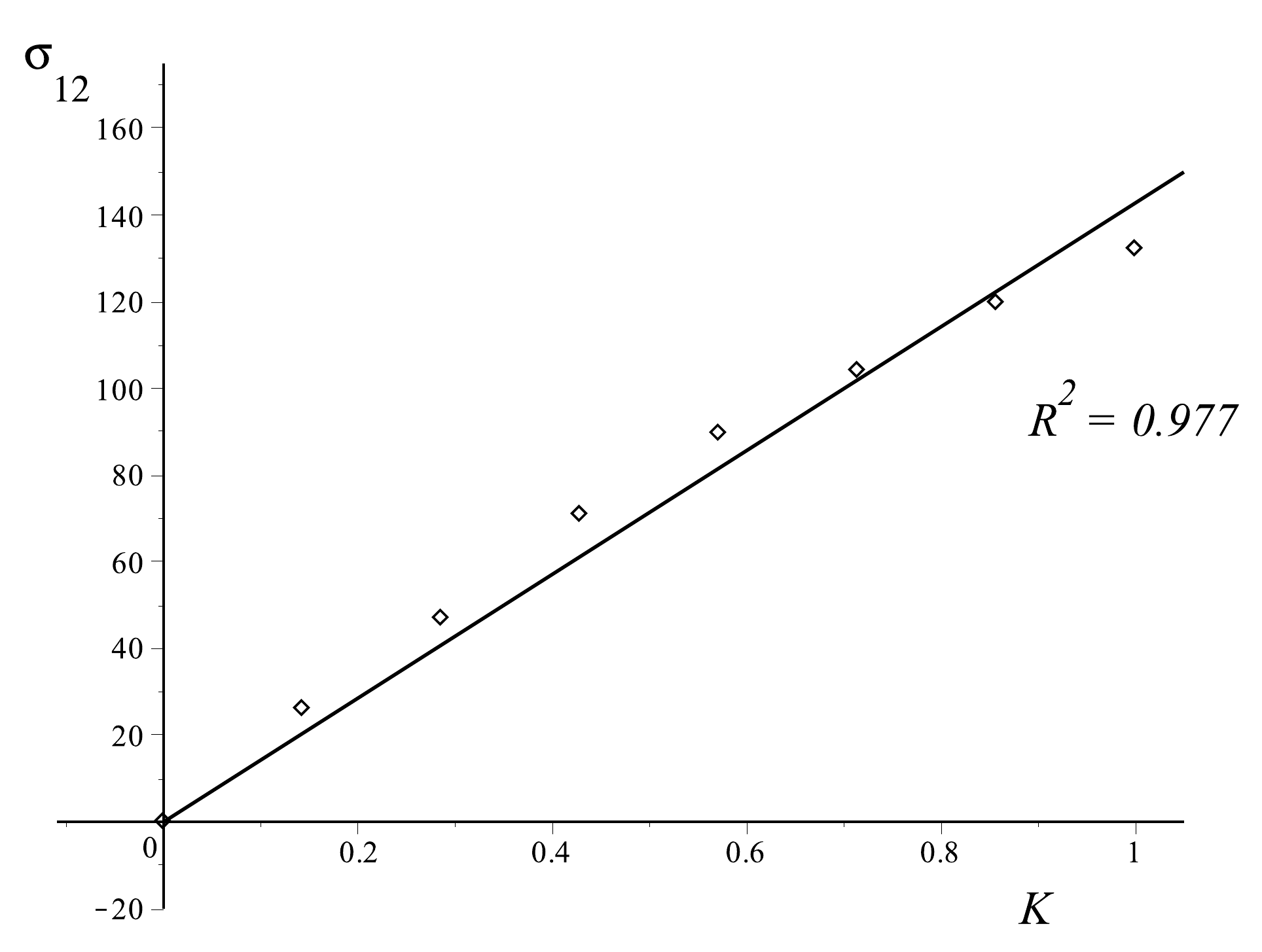, width=0.6\textwidth}}\\
\subfigure{\epsfig{figure=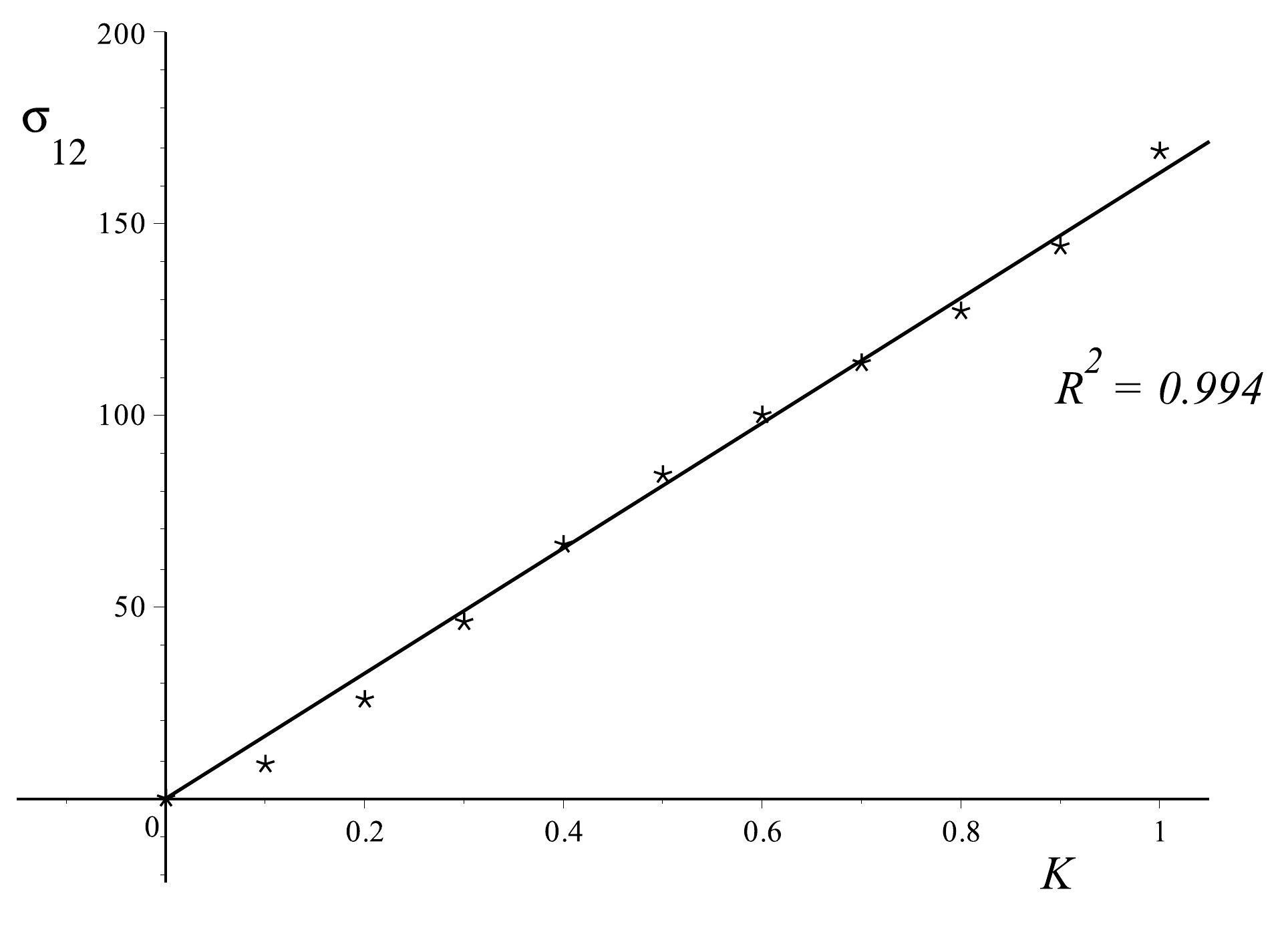, width=0.6\textwidth}}
\caption{
\small
Simple shear experiments on cuboids of Sylgard$\textsuperscript \textregistered$ gel ($19 \times 19 \times 7$ mm) and porcine brain matter  (Sample $A$: $19 \times 19 \times 7$ mm, Sample $B$: $19 \times 19\times3$ mm):  shear Cauchy stress $\sigma_{12}$ (Pa) Vs finite amount of shear $K$ (no dimension). The fitting is made to a linear relationship $\sigma_{12} = \mu K$ (see $R^2$ for a measure of the goodness of fit), showing that these samples of soft matter behave essentially as Mooney-Rivlin materials.
}
\label{fig:shear-stress}
\end{figure}

 A linear $\sigma_{12}-K$ relationship is exhibited for the three samples tested in simple shear test as shown in Fig.\ref{fig:shear-stress}. The linearity is verified for the Sylgard gel, with a coefficient of determination $R^2=0.990$, and for the brain samples, with coefficients of determination $R^2=0.977$ for Sample $A$ and $R^2=0.994$ for Sample $B$. 

From a least-square optimisation, we find that the slope $2(C_1+C_2)$ of the lines is 5060 Pa for the Sylgard gel, 143 Pa for brain Sample $A$, and 163 Pa for brain Sample $B$. 

\subsection{Normal stress}

When the simple shear test was done using the pierced  platen, we recorded that the circular area exposed by the hole tended to bulge outwards, indicating that in the absence of the platen the sample would expand in simple shear. 

\begin{figure}
\subfigure{\reflectbox{ \includegraphics[width=0.50\columnwidth] {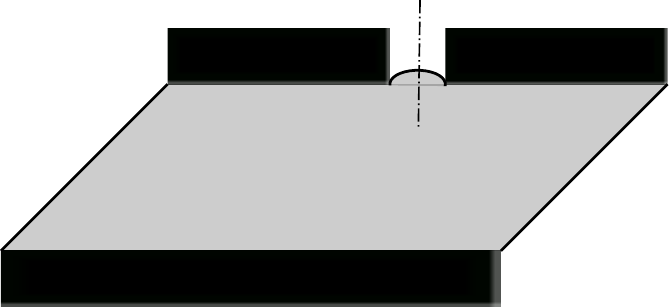} }}
\subfigure{\reflectbox{ \includegraphics[width=0.50\columnwidth] {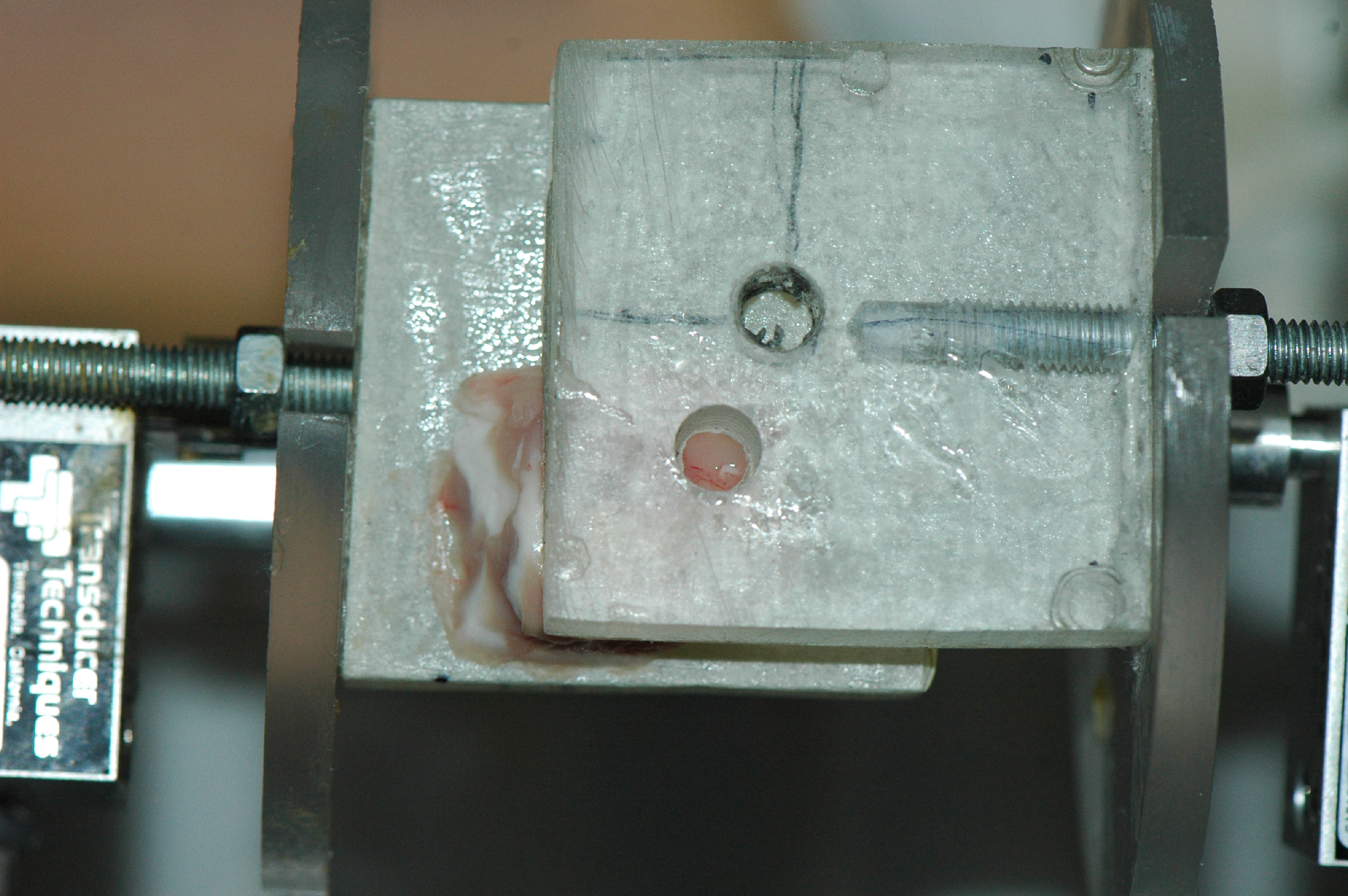} }}
\caption{
\small
Poynting effect for the large ($K=1$, angle $=45^\circ$) simple shear of porcine brain matter  ($19 \times 19 \times 3$ mm): without the platens, the sample would expand in the 2-direction. This tendency is revealed by drilling a hole in one platen and observing from above the drilled platen that the exposed area bulges outwards
(It is clearly visible to the naked eye, but difficult to capture in the photograph).
}
\label{fig:poynting}
\end{figure}

\begin{figure}
\begin{center}
\epsfig{figure=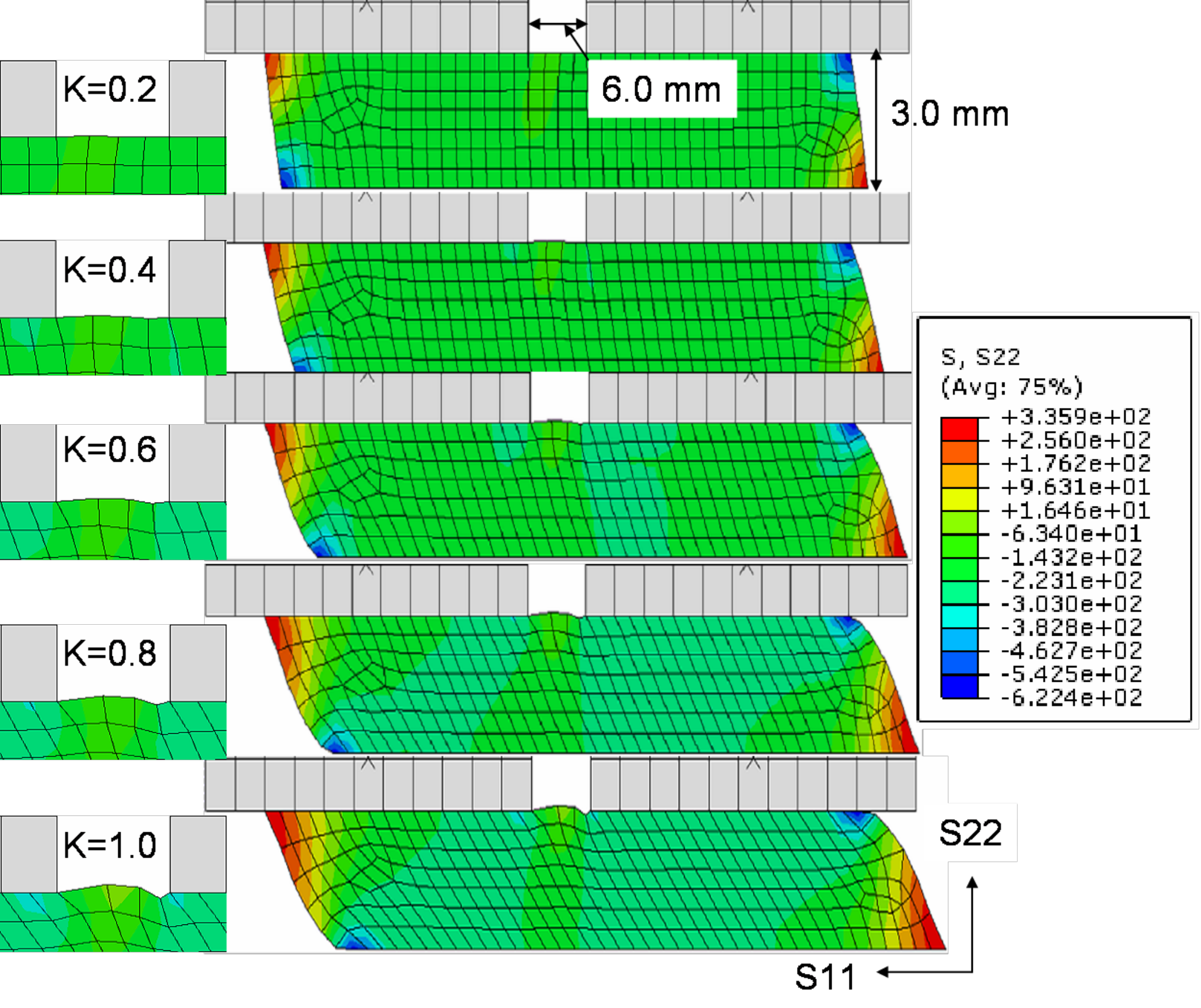, width=0.6\textwidth}
\end{center}
\caption{
\small
Finite Element simulation of simple shear experiment conducted in Fig.\ref{fig:poynting}.
The scale on the right indicates the normal stress magnitude $S_{22} = \sigma_{22}$ at $K=1$. 
The simulation confirms that outside localized edge areas, the stress distribution is largely homogeneous throughout the sample, and that the Poynting effect will make the material bulge out in the absence of a normal compressive force.}
\label{fig:abaqus}
\end{figure}

Several measurements ($n=$4) confirmed the tendency for the brain matter to expand in simple shear, see Fig.\ref{fig:poynting}.  The results of the numerical simulation displayed in Fig.\ref{fig:abaqus} repeats this observation of a bulging out for simple shear, i.e. of a positive Poynting effect.

\section{Discussion}
 
Let us start by recalling that in simple shear, $W=W(K^2)$ so that $\sigma_{12}$ given by \eqref{sigma}  is \emph{always an odd function of $K$}. 
This is in line with the physics of the deformation, as the shear stress necessary to shear the block by amount $K$ is the opposite of that necessary to shear it by an amount $-K$. 
In particular it follows that $\sigma_{12}$ cannot be of the form $aK^b$ where $a$, $b$ are curve-fitting constants as in \cite{DoMe97}.

The proof that \emph{brain matter behaves as a Mooney-Rivlin material} as long as stretches are below $60\%$ is reached here through the verification of the linearity between $\sigma_{12}$ and  $K$. 
This result is not that surprising, because there exist a well-known \cite{Rivl49, GoVD08} connection between the Mooney-Rivlin strain energy density and that of the most general nonlinear, third-order elasticity, isotropic, incompressible material,
\be
W = \mu \, \text{tr}\left(\vec{E}^2\right) + \dfrac{A}{3}\text{tr}\left(\vec{E}^3\right),
\label{TOE}
\en
where $\mu$ is the shear modulus, $A$ is a third-order elastic constant, and $\vec E$ is the Green strain. 
At the same order of approximation, \eqref{MR} and \eqref{TOE} are equivalent, and the constants are connected through \cite{DeGM10}
\be \label{connection}
\mu = 2(C_1+C_2), \qquad
A = -8(C_1+2C_2).
\en

Note that shear force data only provides a value for $\mu=2(C_1 + C_2)$, and normal force data or another testing protocol are required to complete its material characterization.  

The \emph{normal forces} at play during simple shear have been checked qualitatively in this study, through experiments and numerical simulations. Recalling that in linear elasticity, $\sigma_{11}=\sigma_{22}=0$, then the observation of non-zero normal forces for tests confirms that the behaviour of the brain matter is indeed \emph{nonlinear}. This is the well-known (positive) Poynting effect. 
The presence of this  phenomenon also allows us to conclude that brain matter does not belong to the so-called "generalized neo-Hookean class", where $W=W(I_1)$ only, because if it did, then $\sigma_{22}$ would be zero according to \eqref{sigma}$_2$. 
In effect, observing the bulging \emph{out} of the brain matter tells us that $C_2>0$.

The comparison of the experimental slopes (i.e. of the shear moduli $\mu$) between the Sylgard gel and the brain samples show that \emph{brain matter is an extremely soft solid}, at least 30 times less resistant to shearing forces than a silicone gel, when sheared at quasi-static speeds. 
\color{black}We note that at higher strain rates, the shear modulus is much increased, allowing the brain to resist better to shearing tractions during an impact for instance. 
Hence we found \cite{Rashid13} that it was 10 to 20  times higher at strain rates occurring in traumatic brain injury incidents.
Specifically we measured it to be $1.157 \pm 0.25$ kPa, $1.347 \pm 0.19$ kPa, $2.197 \pm 0.225$ kPa and
$2.527 \pm 0.27$ kPa at 30, 60, 90 and 120/s rates, respectively, again for destructive simple shear tests on porcine brains.
\color{black}

The \emph{numerical simulation} of simple shear finally confirms that outside localized edge areas, the stress distribution is largely homogeneous throughout the sample. It makes experimental simple shear a good candidate to extract material properties based on reliable simple analytical models.
As explained earlier, we chose the numerical values of $C_1$ and $C_2$ such that $2(C_1+C_2)=165 $ Pa, in line with the experimental determination of the shear modulus $\mu$.
We picked $C_2$ much larger than $C_1$ to emphasise the visual bulging of the brain matter through the hole.
Lower values of $C_2$ also showed the same effect, as expected.

A shortcoming of our experimental protocol is that it only provides access to the value of $C_1+C_2$ and to the sign of $C_2$, but not to the actual values of $C_1$ and $C_2$ separately. 
To access these values we would need to measure $\sigma_{22}$, which is not a trivial task. 
An alternate, non-destructive, method is to use the large acousto-elastic effect \cite{DeGS10}, which has recently been applied to porcine brains \cite{Jiang}.


\section*{Acknowledgements}


Partial funding by a ``New Foundations'' award from the Irish Research Council is gratefully
acknowledged by the first  author.
We thank M\'elanie Ott\'enio (Universit\'e de Lyon) for her most useful input.


\end{document}